\documentclass[twocolumn]{aastex63}

\usepackage{graphicx}
\usepackage{amsmath}
\usepackage{amssymb}
\usepackage{cases}
\usepackage{mathrsfs}
\usepackage[flushleft]{threeparttable}
\usepackage{amssymb}
\usepackage{pifont}
\usepackage{amsmath}
\usepackage{epstopdf}
\usepackage{xcolor}
\usepackage[all]{hypcap}
\usepackage{mwe,tikz}
\usepackage{bm}
\usepackage{tabularx}
\usepackage{multirow}
\usepackage{booktabs}
\usepackage{epstopdf}
\usepackage{subfigure}
\usepackage{xspace}
\usepackage{rotating}
\usepackage{CJKutf8}

\newcommand{\e}{\text{e}}

\newcommand{\p}{\partial}

\shorttitle{GRB 200826A: a Precursor of a Long GRB}
\shortauthors{Wang et al.}

\begin{document}

\title{GRB 200826A: A Precursor of a Long GRB with Missing Main Emission}
\correspondingauthor{Bin-Bin Zhang}
\email{bbzhang@nju.edu.cn}

\author[0000-0002-9738-1238]{Xiangyu Ivy Wang}
\affiliation{School of Astronomy and Space Science, Nanjing University, Nanjing 210093, China}

\author[0000-0003-4111-5958]{Bin-Bin Zhang}
\affiliation{School of Astronomy and Space Science, Nanjing
University, Nanjing 210093, China}
\affiliation{Key Laboratory of Modern Astronomy and Astrophysics (Nanjing University), Ministry of Education, China}

\author{Wei-Hua Lei}
\affiliation{Department of Astronomy, School of Physics, Huazhong University of Science and Technology, Wuhan 430074, China}

\begin{abstract}

The recently discovered peculiar gamma-ray burst GRB 200826A poses a dilemma for the collapsar model. Although all other characteristics of the burst are consistent with it being a Type II (i.e., collapse of a massive star) event, the observed duration of the event is only approximately 1 s, which is at odds with the predicted allowable timescale range for a collapsar event. To resolve this dilemma, this {\it Letter} proposes that the original burst could be an intrinsically long GRB comprising of a precursor and a main emission (ME) phase. However, the main emission phase is missed due to either precession of the jet or the obstruction of a companion star, leaving only the precursor observed as a short-duration GRB 200826A. Interestingly, we found that the temporal and spectral properties of GRB 200826A broadly resembled those of the bright precursor observed in GRB 160625B. Furthermore, assuming the prototype burst of GRB 200826A is similar to that of GRB 160625B, we found that the observer may indeed miss its main emission because of geometric effects caused either by jet precession or companion-obstruction models. Our approach provides a natural explanation for the GRB 200826A-like bursts and agrees with the rarity of those events.

\end{abstract}


\section{Introduction} \label{sec:intro}
 GRB 200826A challenges the traditional observational criteria used to classify short merger-type and long collapsar-type GRBs \citep{1993ApJ...413L.101K, 2009ApJ...703.1696Z}. With a genius, short duration of $T_{90}$ = $0.96_{-0.07}^{+0.06}$ s \citep{2021NatAs...5..911Z}, GRB 200826A distinguishes itself from all other Type I bursts through all of its other observational properties such as hardness ratio, energy-related correlations, amplitude parameter, spectral lag, and possible supernova association \citep{2021NatAs...5..911Z, 2021NatAs...5..917A, 2021arXiv210503829R}. Mounting efforts have been made to explain this peculiar GRB. For example, 
\cite{2021NatAs...5..917A} claim that it is the shortest collapsar event, and \cite{2021NatAs...5..911Z} additionally 
propose several alternatives such as a binary merger invoking a white dwarf (WD) engine, the ``supranova" scenario, and a newborn magnetar with a heavy baryon-loading wind. Rather than making any modifications to the GRB's central engine, we propose in this {\it Letter} that GRB 200826A was actually the precursor emission of a long GRB whose main emission(ME) was not observed due to geometric effects.

In terms of geometry, there are two possible configurations that might lead to the absence of GRB emission. One possibility is that the jet slips away, in which case the observer is no longer within the jet cone. Such a slip-away effect can be caused by the precession of the jet. In such a scenario, the jet is powered by a hyperaccreting black hole, and the angular momentum of the black hole is misaligned with respect to that of the accretion disk. The tilted disk is subject to the Lense-Thirring (LT) torque. The LT torque together with the viscosity of the disk causes the inner part of the inclined disk to bend toward the equatorial plane of the black hole, while the outer part of the disk maintains its original orbit \citep{1918PhyZ...19..156L, 1975ApJ...195L..65B}. The inner disk would undergo precession. As a result, a precessing disk will induce jet precession \citep{2006A&A...454...11R, 2010A&A...516A..16L, 2013ApJ...762...98L}. 

A second possibility is that the jet is completely blocked during the ME, so the observer only observes the precursor. This may occur if the GRB is in a binary system \citep{2021ApJ...921....2Z}. Such a system is composed of a GRB central engine and a stellar companion. If the observer is on axis and the companion star is located within the jet opening angle\footnote{Here we consider a top-hat jet with $\theta_j\gg 1/\Gamma$ during the prompt emission phase of the GRB, where $\Gamma$ is the Lorentz factor of the ejecta. In this case, only a small fraction of the emitting surface of the jet is observable. This region is centered on the line of sight and has an opening angle of 1/$\Gamma$.} \citep[case I in][]{2021ApJ...921....2Z}, the observed GRB properties are determined by the observation angle (the angle between the jet direction and the observer's line of sight), the Lorentz factor of the jet, and the obstruction by the companion star. If the observer is on axis, and the Lorentz factor is greater than a critical value, the companion star can block the GRB emission so that the on-axis observer completely misses it. For GRB 200826A, we demonstrate in this {\it letter} that the ME can be entirely blocked by using an appropriate Lorentz factor value and geometric configuration.

This {\it Letter} starts by comparing the observed properties of GRB 200826A with the precursor of the typical three-episode GRB 160625B, aiming to find observational evidence of their similarities (\S \ref{sec:comp}). The jet-precession (\S \ref{sec:precessing}) and companion-obstruction (\S \ref{sec:block}) models are then applied to explain the observations of GRB 200826A as well as its possibly missing ME. A brief conclusion and discussion are presented in \S \ref{sec:conclusion}.

\section{GRB 200826A as a Precursor}

\label{sec:comp}

\begin{figure}
 \centering
 \includegraphics[width=0.47\textwidth]{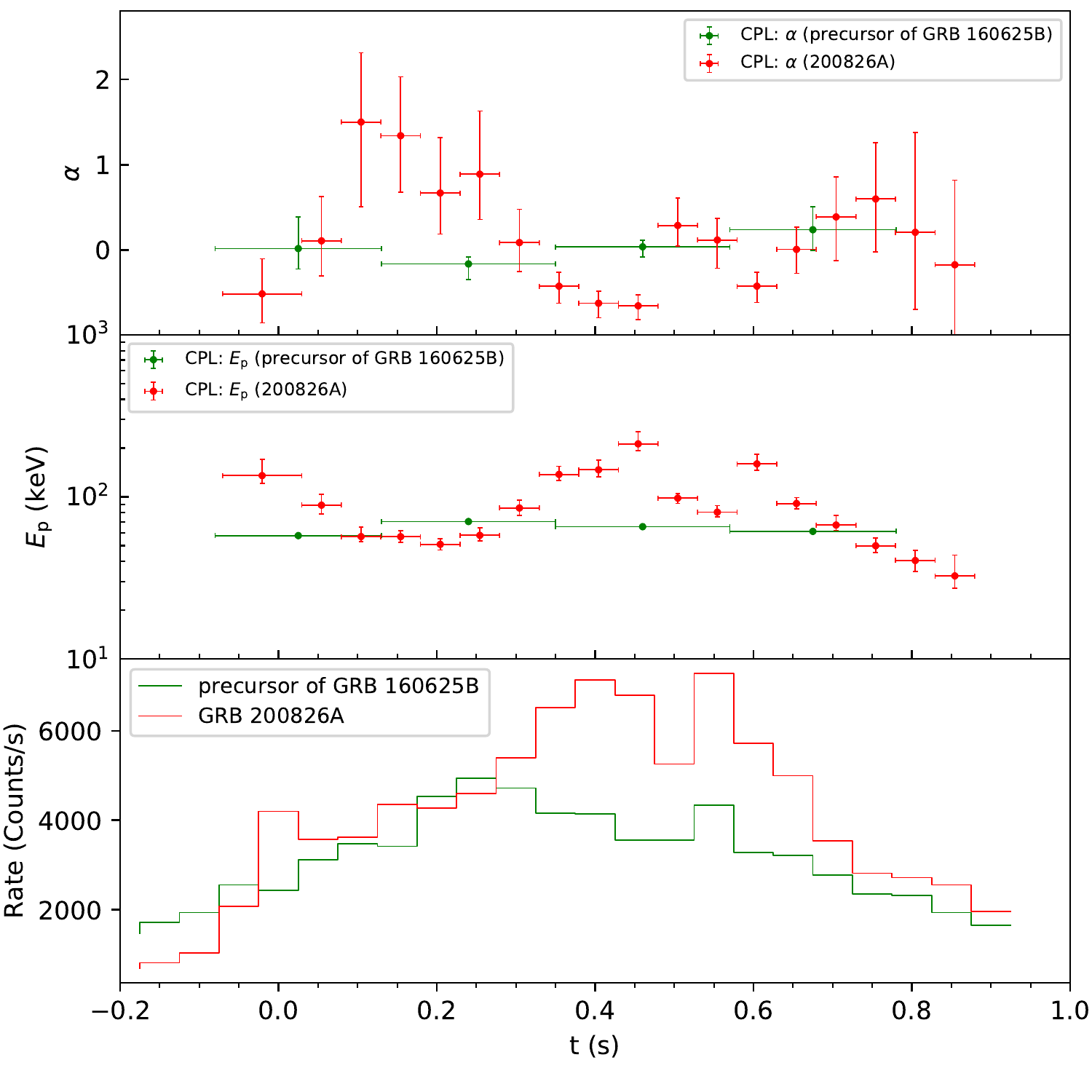}
 \caption{Comparison of the spectral evolution between GRB 200826A and the precursor of GRB 160625B. Both bursts are fitted by the CPL model. The bottom panel shows the comparison of the light curves of the two events. Data are taken from \cite{2018NatAs...2...69Z} and \cite{2021NatAs...5..911Z}.}
 \label{fig:spe_evo}
\end{figure}

\begin{figure}
 \centering
 \includegraphics[width=0.47\textwidth]{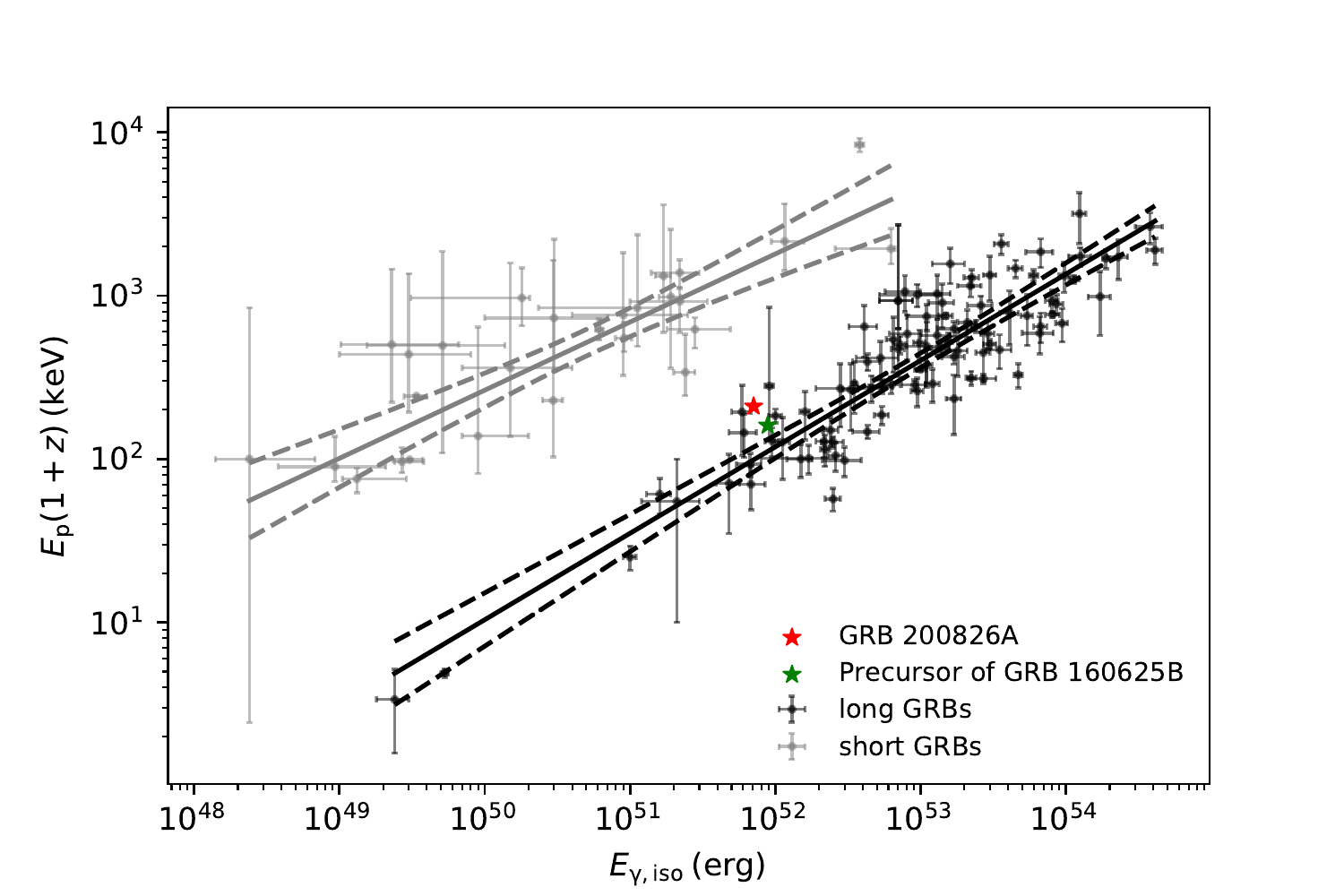}
 \caption{The $E_{\rm p, z}$ vs. $E_{\rm\gamma,iso}$ correlation diagram. The black and gray solid lines are the best linear correlations for long and short GRBs, respectively. GRB 200826A (red star) and the precursor of GRB 160625B (green star) both fall in the long-GRBs region. The samples are from \cite{2002A&A...390...81A} and \cite{2009ApJ...703.1696Z}}.
 \label{fig:Ep_Eiso}
\end{figure}

\begin{figure}
 \centering
 \includegraphics[width=0.47\textwidth]{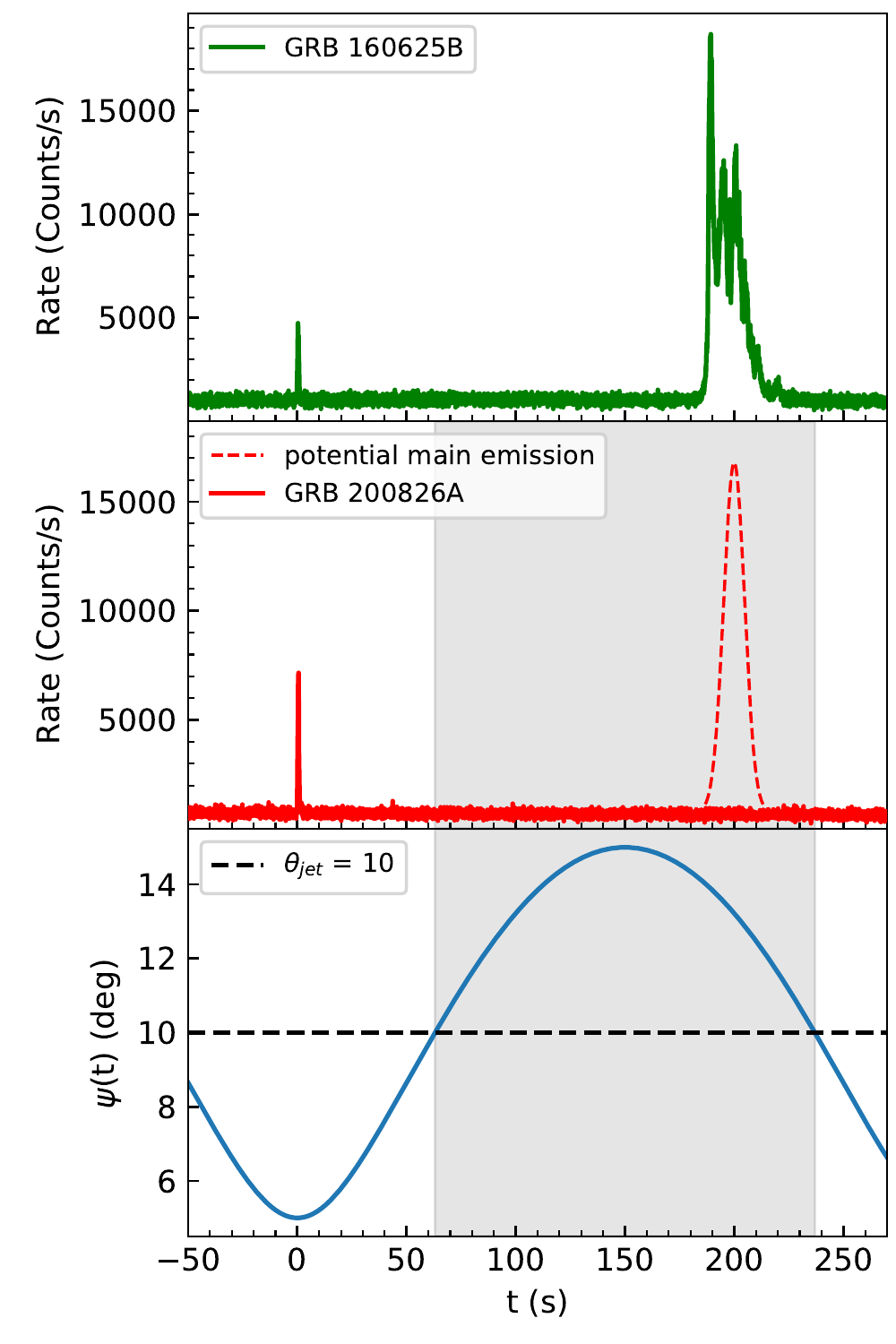}
 \caption{{\it Top:} light curve of GRB 160625B. {\it Middle:} The solid red line is the light curve of GRB 200826A. The dashed red line is the assumed ME of the prototype burst. {\it Bottom:} The solid blue line shows how the viewing angle $\psi(\rm t)$ changes over time, and the dashed black line is the jet-opening angle. The gray shaded area is where the ME can be missed by the observer.}
 \label{fig:lc}
\end{figure}

GRB 200826A is considered a precursor of a long GRB due to the following facts:
\begin{enumerate}
 \item Temporal and spectral properties consistent with the precursors of other GRBs. GRB precursors are always characterized by a short duration and thermal emission \citep{2007MNRAS.380..621L}. Indeed, the duration of GRB 200826A is only 1 s, similar to the typical duration of precursors in long GRBs \citep[e.g., GRB 160625B;][]{2018NatAs...2...69Z}. Moreover, the spectral evolution of GRB 200826A (Figure \ref{fig:spe_evo}) shows that its low-energy indexes, $\alpha$, are greater than zero in about two-third of the time slices, which are consistent with being a thermal origin. 
 
 \item Location on the $E_{\rm p} - E_{\rm iso}$ diagram consistent with those of the precursors in other GRBs. The burst is located at the long (Type II) GRB track on the $E_{\rm p} - E_{\rm iso}$ diagram, which is similar to the precursors of other long GRBs. As an example, we plot both GRB 200826A and the precursor of GRB 160625B, a typical three-episode-long GRB with a significant precursor emission, on the $E_{\rm p} - E_{\rm iso}$ diagram in Figure \ref{fig:Ep_Eiso}. One can see that the two events both reside on the Type II GRB track. 
 
 \item Almost identical to the precursor of GRB 160625B. In Table \ref{tab:compara_para} and Figure 1, we compare the temporal and spectral properties between GRB 200826A and the precursor phase of GRB 160625B. The two events display striking similarities in the following aspects: (1) similar duration. Both of their $T_{90}$ values are roughly equal to 1 s. (2) Similarly strong signal-to-noise ratios. As shown in Table 1, the f-parameter values, which measure the ``tip-of-iceberg" effect of a GRB \citep{2014MNRAS.442.1922L}, of the two GRBs are both above 3, indicating strong signals above the background (Figure \ref{fig:lc}). (3) Similar temporal and spectral evolution patterns. As shown in Figure \ref{fig:spe_evo}, the light curves of both events are single-pulse shaped. Although GRB 200826A has a stronger spectral evolution, its spectral parameters (such as the low-energy index, $\alpha$, and spectral peak energy, $E_{\rm P}$) are overall in a range consistent with those of the precursor of GRB 160625B. Both events exhibit strong thermal-like spectral features in $>80$\% time slices. 

\end{enumerate}

In summary, GRB 200826A is fully consistent with being a precursor of a long-GRB event. The question is how the ME of such a long GRB can be missed by an observer.

\begin{table*}[]
 \centering
 \begin{tabular}{c|cc}
 \hline
 \hline
 & GRB 200826A & Precursor of GRB 160625B\\
 \hline
 z & 0.7481 & 1.406 \\
 $T_{90}$ (s) & $0.96_{-0.07}^{+0.06}$ & $0.84_{-0.01}^{+0.03}$ \\
 f-parameter & $7.58\pm1.23$ & $3.42\pm0.14$\\
 Peak energy (keV) & $120.29_{-3.67}^{+3.93}$ & $66.8_{-1.8}^{+1.8}$ \\
 Peak flux ($10^{-6}\ {\rm erg} {\rm cm^{-2}\ s^{-1}}$) & $9.11_{-1.17}^{+1.47}$ & $2.42_{-0.11}^{+0.11}$\\
 Fluence ($10^{-6}\ {\rm erg} {\rm cm^{-2}}$) & $4.85\pm 0.19$ & $1.75\pm 0.05$\\
 Isotropic energy ($10^{51}\ {\rm erg}$) & $7.09\pm 0.28$ & $8.86\pm 0.24$\\
 \hline
 \end{tabular}
 \caption{Properties of GRB 200826A and the Precursor of GRB 160625B \citep{2021NatAs...5..911Z, 2018NatAs...2...69Z}.}
 \label{tab:compara_para}
\end{table*}

\section{How can We miss the main emission?}

To miss the ME, either the jet had to slip away within a certain time frame or the jet had to be blocked during the ME. The two scenarios correspond to the following two different physical pictures.

\label{sec:jet_precessing}
\subsection{Jet Precession}
\label{sec:precessing}

\begin{figure}
 \centering
 \includegraphics[width=0.47\textwidth]{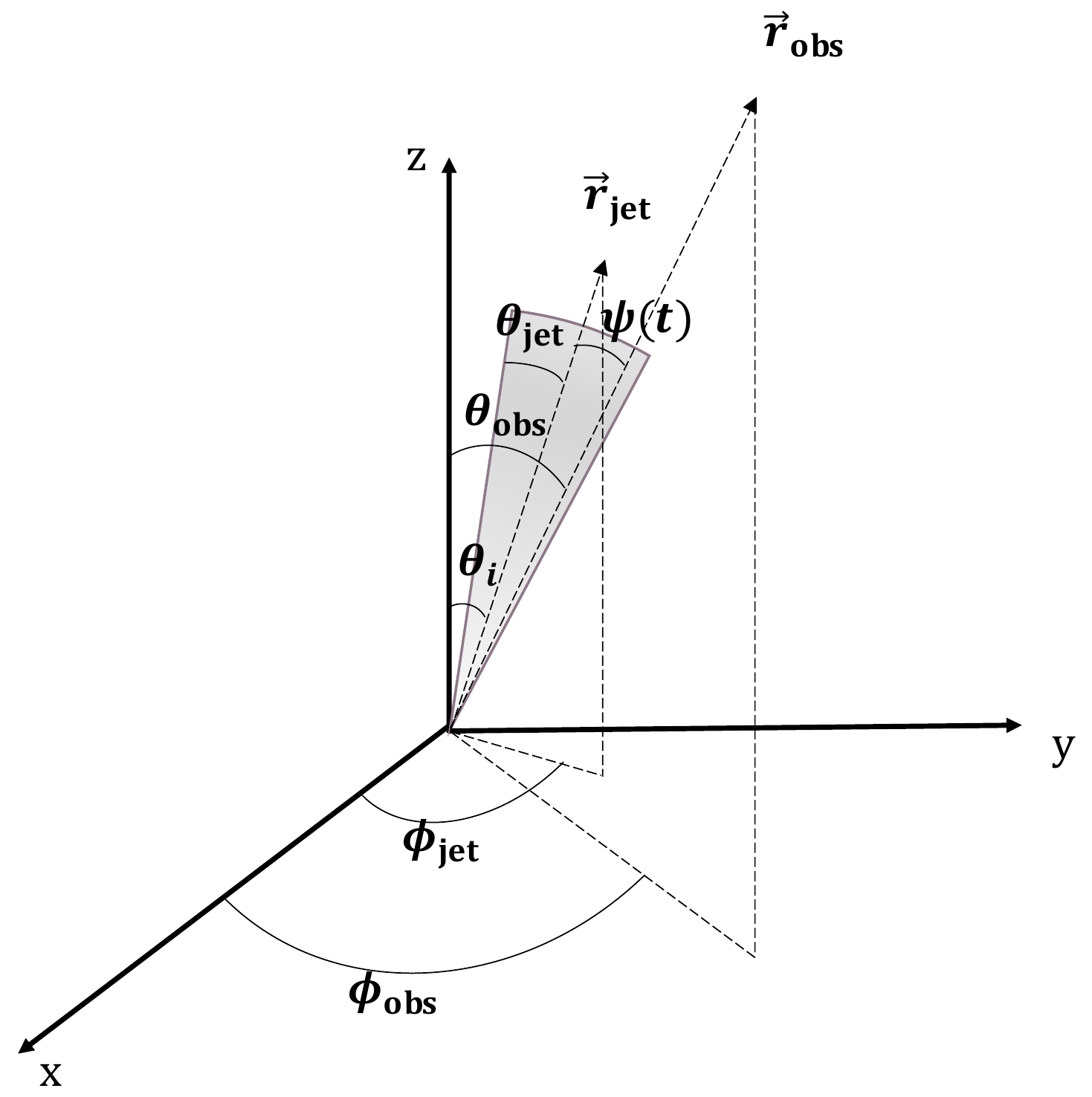}
 \caption{The schematic sketch of the precession jet model.}
 \label{fig:model}
\end{figure}

\begin{figure}
 \centering
 \includegraphics[width=0.47\textwidth]{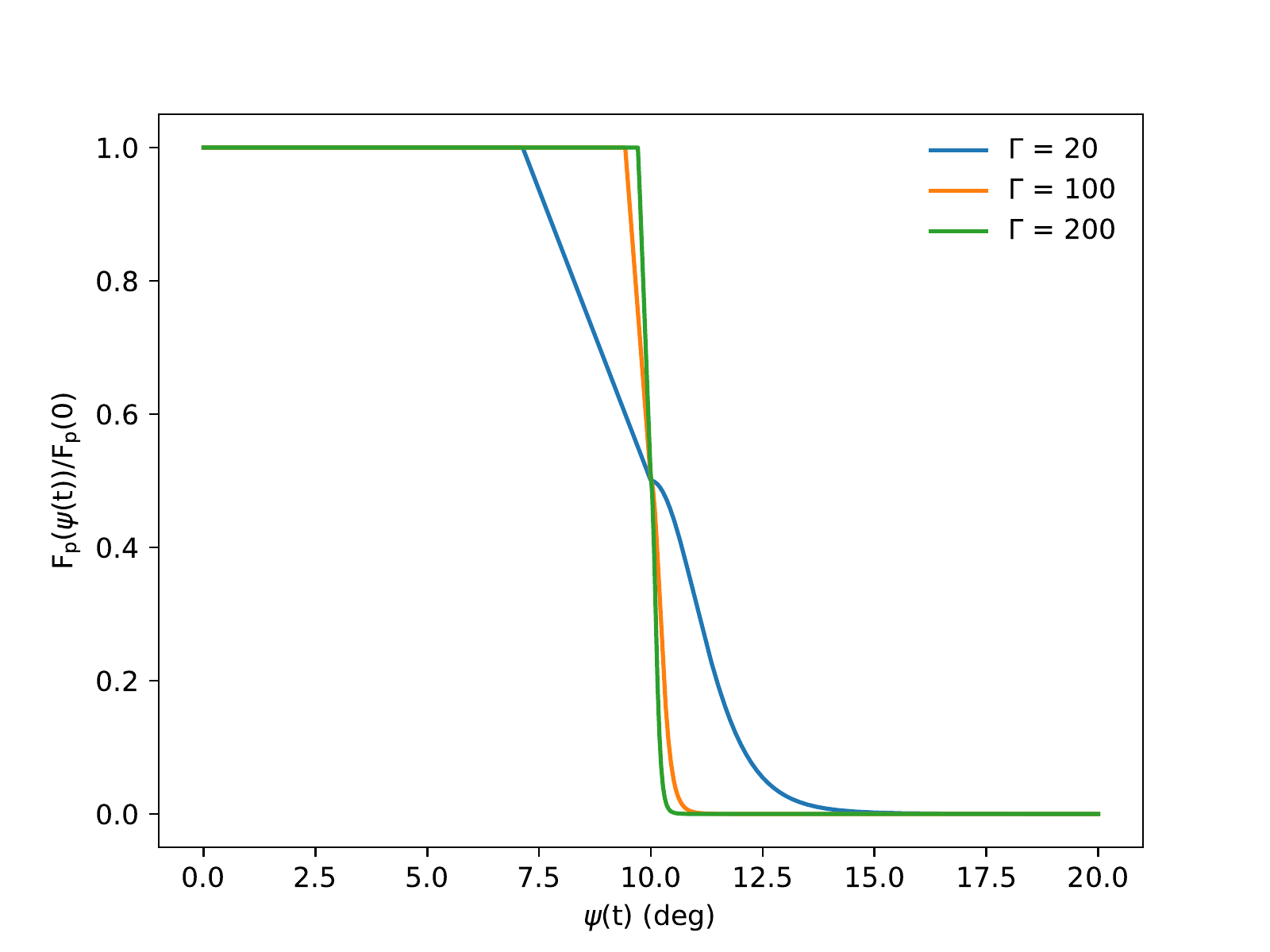}
 \caption{The blue, orange, and green lines represent the evolution of the observed peak flux with the viewing angle when the Lorentz factor $\Gamma$ = 20, 100, and 200, respectively.}
 \label{fig:lorentz}
\end{figure}

Jet precession has long been proposed \citep[e.g., see ][]{2010A&A...516A..16L} for GRBs whose central engine consists of a rapidly hyperaccreting black hole. For a long GRB, the anisotropic explosions of its progenitor star lead to the misalignment between the disk and angular momentum of the black hole. In such a scenario, the Bardeen–Petterson (BP) effect \citep{1975ApJ...195L..65B} tends to align the inner part of the inclined disk with the equator of the black hole while the outer part of the disk maintains its original orbit. The outer disk will lead to the precession of the inner disk and the black hole \citep{1980ApJ...238L.129S, 2010A&A...516A..16L, 2012ApJ...752...31S} due to the LT effect, which is known as disk-driven precession. The jet will precess because its direction is determined by the spin axis of the black hole, and its precession period is the same as the LT precession period.

The jet precession model has been utilized to interpret various shapes of GRB light curves \citep[e.g.,][]{1999ApJ...520..666P, 2007A&A...468..563L} and their the complex temporal and spectral evolution \citep{2010A&A...516A..16L}. There is a possibility that, in some cases, the ME might be missed as a result of the jet precession. Here we will show that GRB 200826A may be such a case.

As schematized in Figure \ref{fig:model}, we assume the jet is conical and no moving material is outside the cone. \cite{2016MNRAS.461.3607S} studied the effect of viewing angle $\psi(\rm t)$ on GRB peak flux and showed that the peak flux can be written as a function of $\psi(\rm t)$:
\begin{equation}
\label{eq:obs_flux1}
 \rm F_{p}(\psi(\rm t))/F_{p}(0) = 
	\left\{
		\begin{array}{lll}
			1, & \theta_{\rm jet} \ge \theta_{\rm jet}^{*},\\
			\frac{1-\Gamma(\psi(\rm t) - \theta_{\rm jet})}{2}, & \theta_{\rm jet}^{*} < \psi(\rm t) \le \theta_{\rm jet},\\
			\frac{1}{2}\Big(\frac{D}{(1+\beta)\Gamma}\Big)^{\big(4-\sqrt{2}\theta_{\rm jet}^{1/3}\big)}, & \psi(\rm t) > \theta_{\rm jet},
		\end{array}
	\right.
\end{equation}
where $\theta_{\rm jet}^{*}$ = $\theta_{\rm jet}-\frac{1}{\Gamma}$, $\beta$ = $\sqrt{1-\Gamma^{-2}}$, $\rm F_{p}(0)$ is the observed peak flux when the line of sight is centered on the jet axis, and D is the Doppler factor defined as $\rm D=\frac{1}{\Gamma(1-\beta\cos{\psi(t))}}$. 

Figure \ref{fig:lorentz} shows the observed peak flux changes over the viewing angle $\psi(\rm t)$ with $\theta_{\rm jet}$ = 10$^{\circ}$. One can see that the larger the Lorentz factor, the sharper the peak flux drops as the line of sight in the direction of $\vec{r}_{\rm obs}$ moves away from $\theta_{\rm jet}^{*}$. For a jet with a Lorentz factor of $\Gamma>$ 200 \citep[e.g., $\Gamma \ge$ 214 for the precursor of GRB 160625B;][]{2018NatAs...2...69Z}, we can assume that no flux would be received by the observer once $\psi(\rm t)$ exceeds $\theta_{\rm jet}$, and the observed flux can be written as a function of $\psi(\rm t)$:
\begin{equation}
\label{eq:obs_flux2}
 \rm F(\psi(\rm t)) = 
	\left\{
		\begin{array}{lll}
			\rm F(0), & \theta_{\rm jet} > \psi(\rm t),\\
			0, & \theta_{\rm jet} < \psi(\rm t).
		\end{array}
	\right. 
\end{equation}

Following \cite{2014ApJ...781L..19H, 2014MNRAS.441.2375H}, the precession period can be expressed as
\begin{align}
 \label{eq:period}
 \tau = 2793a_{\bullet}^{\frac{17}{13}}\left(\frac{M_{\rm BH}}{\rm M_{\odot}}\right)^{\frac{7}{13}}\left(\frac{\dot{M}}{\rm M_{\odot} \rm s^{-1}}\right)^{-\frac{30}{13}}\alpha^{\frac{36}{13}} {\mathrm{\ s}},
\end{align}
where $a_{\bullet}$ is the spin parameter, $M_{\rm BH}$ is the mass of the rotating black hole, $\dot{M}$ is the accretion rate in the units of $\rm M_\odot s^{-1}$, and $\alpha$ is the viscosity parameter of the accreting disk. In the rest frame of the observer, the precession period is $\tau'$ = (1+z)$\tau$. For a standard collapsar model \citep{1999ApJ...524..262M}, the 14$\rm M_{\odot}$ helium core of a 35$\rm M_{\odot}$ main-sequence star would collapse to form a 2 to $\sim$3$\rm M_{\odot}$ black hole. Inserting $a_{\bullet}$ = 0.9, $M_{\rm BH}$ = 3$\rm M_{\odot}$, $\dot{M}$ = 0.1$\rm M_{\odot}$ $\rm s^{-1}$, and $\alpha$ = [0.01, 0.1] into equation \ref{eq:period}, we get $\tau'$ = [4.52, 2655.67] s. Here, we take $\tau'$ = 300 s in the following analysis.

For a conical jet, the angle $\psi(\rm t)$ between the observer and the jet axis can be written as:
\begin{align}
 \label{eq:psi}
 \cos(\psi(\rm t)) = \cos\theta_{\rm obs}\cos\theta_{\rm i} + \sin\theta_{\rm obs} \sin\theta_{\rm i} \cos\beta,
\end{align}
where $\beta$ = $\frac{2\pi}{\tau}t$ + $\beta_{\rm 0}$, and $\beta_{\rm 0}$ = $\phi_{\rm obs}$ - $\phi_{\rm jet}$. Taking $\theta_{\rm obs}$ = 10$^{\circ}$, $\theta_{\rm i}$ = 5$^{\circ}$, $\beta_{0}$ = 0, and $\theta_{\rm jet}$ = 10$^{\circ}$, the observational angle $\psi(\rm t)$ as a function of time is plotted in the lower panel of Figure \ref{fig:lc}. \textcolor{black}{According to Eq. \ref{eq:obs_flux2},} when $\psi(\rm t)$ is larger than the jet-opening angle (the black dashed line), the GRB flux will be missed by the observer. The gray shaded area in Figure \ref{fig:lc} marks the time range during which the GRB emission can be missed. Using the assumption that the prototype GRB 200826A is a GRB 160625B-like event that has an ME of $40$ s duration around $t=200$ s, one can find that the ME can be completely missed due to the precession effect.

\subsection{Companion star Obstruction}
\label{sec:block}

\begin{figure}
 \centering
 \includegraphics[width=0.45\textwidth]{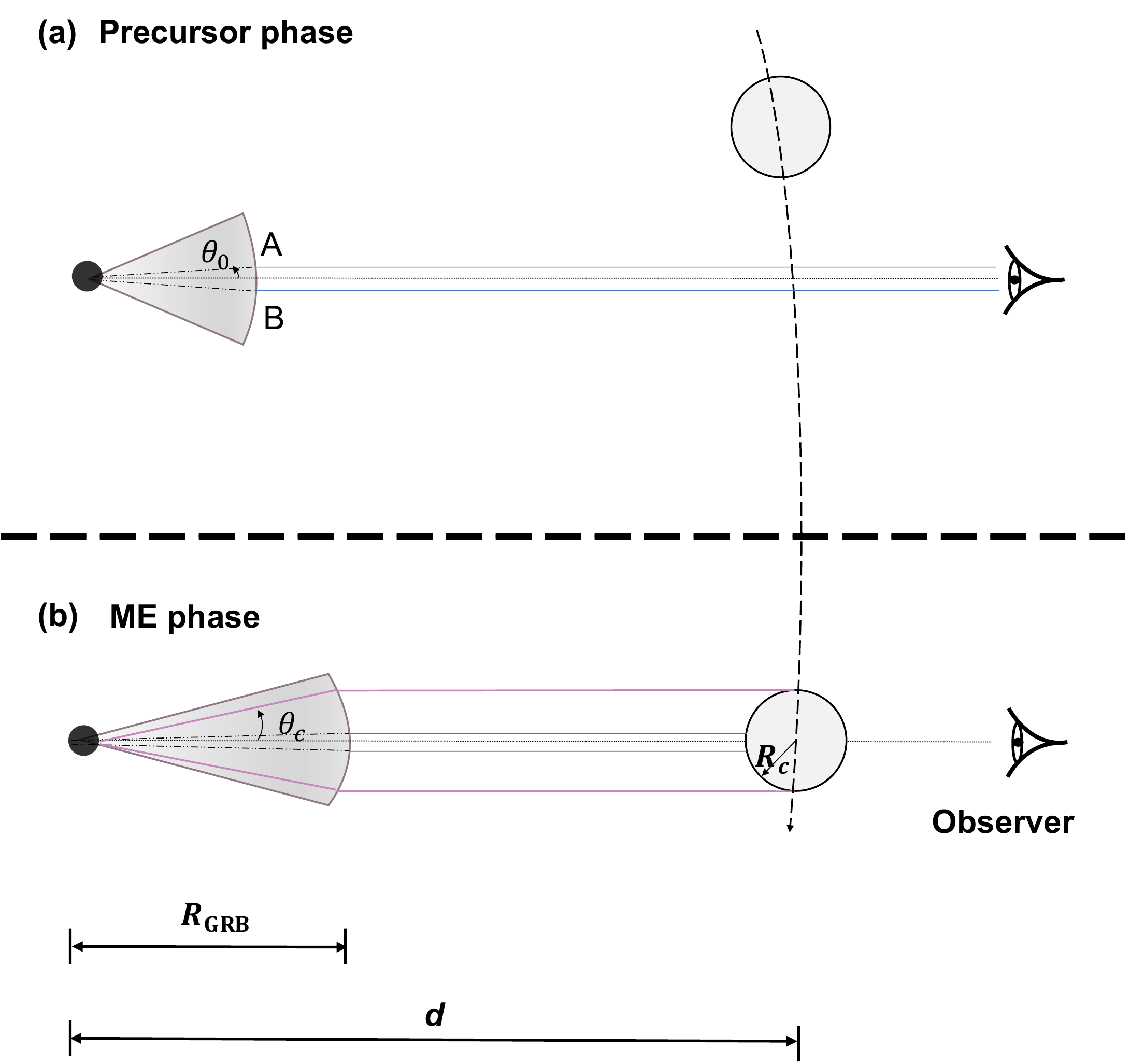}
 \caption{The schematic sketch of the companion-obstruction model. See also \cite{2021ApJ...921....2Z}. (a) The configuration of the system during the precursor phase. (b) The configuration for the ME phase.}
 \label{fig:blokage}
\end{figure}

If a GRB is accompanied by a stellar companion with a radius of $R_{\rm c}$, which happens to be located inside the jet-opening angle, the GRB emission with a radius of $R_{\rm GRB}$ can be partially or fully blocked \citep[][]{2021ApJ...921....2Z}. In such a scenario, the absence of the ME can be explained by the full obstruction by the companion star, as illustrated in Figure \ref{fig:blokage}. Firstly, the precursor is not blocked before the companion star enters within a certain solid angle (calculated below) along the line of sight, allowing the observer to receive the precursor emission as usual (Figure 6a). Then, due to the orbital motion, if the companion star moves into the line of sight and meets the requirements for a full-obstruction condition, the observer can entirely miss the ME (Figure 6b). Assuming the radiation is from the radially relativistically expanding surface, the radiation of each point of the emission surface will beam into a conical angle of radius 1/$\Gamma$ in the direction of its velocity. Consequently, the observer will only receive photons within the 1/$\Gamma$ cone of a conical jet along the line of sight (between A and B in Figure \ref{fig:blokage}). We assume a typical top-hat jet with an opening angle of $\theta_{\rm jet}\sim$ a few degrees, much larger than the beaming angle of $1/\Gamma$ as the Lorentz factor may reach a several hundred during the prompt-emission phase. The full-obstruction condition can be expressed numerically as follows.

\begin{enumerate}
 \item The observer can only receive the photons within the $\theta_{0}$ cone. $\theta_{0}$ is defined as $\theta_{0}$ = $\frac{1}{\Gamma}$ with $\Gamma$ being the Lorentz factor of the ejecta. The maximal obstruction angle by the companion star \citep[i.e., the opening angle of the blocked emission surface with respect to the GRB central engine;][]{2021ApJ...921....2Z}, $\theta_c=\frac{R_{\rm c}}{R_{\rm GRB}}$, must exceed the maximal observed cone angle $\theta_0$;

 \item During the time gap between the precursor and ME, the motion angle of the companion star must exceed $2\theta_{0}$ in order to occult the ME;
 \item During the ME phase, the motion angle of the companion star could not exceed $2\theta_{c} - 2\theta_{\rm 0}$. Otherwise, the observer would receive the flux of the ME at its late stage.
\end{enumerate}

These conditions can be written as

% \begin{subnumcases}{}
% \theta_{c} \ge \theta_{0} \label{eq:a};\\
% \Omega t_{w} \ge \frac{2\theta_{0}\pi}{180}\label{eq:b};\\
% \Omega t_{m} \le \frac{(2\theta_{c} - 2\theta_{0})\pi}{180} \label{eq:c},
% \end{subnumcases}

\begin{subnumcases}{}
\theta_{c} \ge \theta_{0} \label{eq:a};\\
\Omega t_{w} \ge 2\theta_{0}\label{eq:b};\\
\Omega t_{m} \le 2\theta_{c} - 2\theta_{0} \label{eq:c},
 \end{subnumcases}
where $\Omega$ = $\sqrt{\frac{GM_{\rm total}}{d^3}}$ is the angular velocity of the companion star, $M_{\rm total}$ is the total mass, $d$ is the distance between the central engine and the stellar companion, $t_{m}$ is the duration of the ME, and $t_{w}$ is the waiting time between the precursor and the ME. In our analysis, $t_{m}$ = 35 s and $t_{w}$ = 180 s \citep{2018NatAs...2...69Z} are adapted in accordance with the those of GRB 160625B.

Assuming a typical parameter set with $M_{\rm total}$ = 50$\rm M_{\odot}$, $\Gamma$ = 800, $R_{\rm GRB}$ = $10^{11}$ cm, \citep[i.e., a typical photoshphere radius, however, c.f.][]{2018NatAs...2...69Z}, and $R_{\rm c}$ = 0.02$\rm R_{\odot}$ (1.39 $\times 10^{9}$ cm), $d$ can be constrained at $d \le$ 3.25 $\times 10^{12}$ cm according to Eq. \ref{eq:b}. 
\ \\

\section{Summary and Discussions}
\label{sec:conclusion}

In this paper, we suggested that the Type II short GRB 200826A may actually be a precursor of a long GRB whose main emission was missed by the observer. The absence of the main emission is further explained by geometrical models that invoke either the precession of the jet or obstruction by the companion star. By assuming that GRB 160625B serves as a prototype of GRB 200826A, we were able to successfully apply these two models and reproduce the GRB 200826A event by omitting the ME of the prototype burst. Even though both models provide acceptable fits, the companion-obstruction model requires more fine-tuning of the parameters and even coincidental alignment between the GRB and the companion star, which, on the other hand, is in agreement with the rarity of the event. Nevertheless, our results shed some alternative light on how to explain the GRB 200826A-like events. Future observations of similar events will be helpful to test the hypothesis proposed in this paper.

\section*{acknowledgements}

We thank Z.-C. Zou and X.-H. Zhao for helpful discussions on the paper. B.B.Z acknowledges support by the National Key Research and Development Programs of China (2018YFA0404204), the National Natural Science Foundation of China (Grant Nos. 11833003, U2038105, 12121003, U2038107, U1931203), the science research grants from the China Manned Space Project with NO.CMS-CSST-2021-B11, and the Program for Innovative Talents, Entrepreneur in Jiangsu. We acknowledge the use of public data from the Fermi Science Support Center (FSSC).

%\bibliography{ms.bib}

\end{document}